\documentstyle [12pt]{article}
\begin{document}
\pagestyle{empty}
\vfill
\eject
\begin{flushright}
SUNY-NTG-94-35
\end{flushright}

\vskip 2.0cm
\centerline{\bf General Correlation Functions in the Schwinger Model}
\centerline{\bf at Zero and Finite Temperature}
\vskip 2.0 cm

\centerline{James V. Steele, Ajay Subramanian, and Ismail Zahed}
\vskip .2cm
\centerline{Department of Physics}
\centerline{SUNY, Stony Brook, New York 11794-3800}
\vskip 2cm

\centerline{\bf Abstract}
\vskip 3mm
\noindent
The general correlations between massless fermions are calculated in
the Schwinger model at arbitrary temperature. The zero temperature
calculations on the plane are reviewed and clarified. Then the finite
temperature fermionic Green's function is computed and the results on
the torus are compared to those on the plane. It is concluded that a
simpler way to calculate the finite temperature results is to
associate certain terms in the zero temperature structure with their
finite temperature counterparts.

\vfill
\noindent
\begin{flushleft}
SUNY-NTG-94-35\\
\today
\end{flushleft}
\eject
\pagestyle{plain}
\setcounter{page}{1}
{\bf 1. Motivation}
\bigskip

A testimony to the complexity of QCD is in the many outstanding
problems of the theory that are still not solved. For example, it is
not fundamentally understood how quarks are confined or chiral
symmetry is broken, though nature seems to adhere to these
properties. Many have studied the Schwinger model (QED in two
Euclidean dimensions with a massless fermion)
\cite{corrf,smilga1,smail} in the hopes of gaining intuition for
tackling various such problems.

Schwinger \cite{schwing} proposed this model to show that gauge
invariance of a vector field need not imply the existence of a
massless particle.  The gauge field phase shifts the right and left
handed massless fermions with respect to one another causing a
breakdown of chiral symmetry.  This leads to an anomaly which takes
form as the generation of a photon mass ($m=g/\sqrt{\pi}$). The vacuum
has a non-trivial topology analogous to the case in QCD.  This theory
also has charge confinement because of Gauss' law in one space
dimension.

Since the Schwinger model is exactly solvable, having a form for the
general $n$-point correlation function would be useful in many
calculations.  The result has been approached on the plane \cite{bc},
on the sphere \cite{jay}, and on the torus \cite{sw}.  Compact
surfaces are preferred in calculations because the infinities are kept
under control. Fermions on the sphere are difficult to define and for
finite temperature considerations the boundary conditions of the torus
are more natural. Therefore, work on the torus is desirable.

Interest in finite temperature calculations in field theories have led
to the evaluation of two- and four-point correlation functions
\cite{zahed}.  In this paper, we will provide explicit calculations for
a six-point correlation function at finite temperature. When combined
with the previous results, the present result yields the generic form
for the $n$-point correlation function for the Schwinger model at finite
temperature. Exact finite temperature calculations of correlation
function in field theory are of paramount importance for the
understanding of finite temperature spectral functions, reflecting on
the effects of matter on the fundamental degrees of freedom. Real- and
Euclidean-time correlators offer important information on fundamental
issues related to symmetry restoration.

Our paper is organized as follows: in section 2, we give a brief
analysis of the two-dimensional Euclidean Dirac operator in the
plane. In section 3, we consider $n$-point scalar correlation
functions in the two-dimensional Euclidean plane. In section 4, the
same analysis is extended to the Euclidean two-dimensional torus, and
the three-point scalar correlator is explicitly worked out. The result
is extended to the $n$-point scalar correlator at finite temperature. A
brief discussion is given in section 5. Some useful algebra can be
found in the four appendices.

\bigskip
{\bf 2. Preparation}
\bigskip

The Dirac operator may be diagonalized with a complete set of
orthonormal eigenfunctions:
$i\raise.15ex\hbox{$/$}\kern-.53em\hbox{$D$} \psi_n = i\gamma_{\mu}
(\partial_{\mu} - igA_{\mu}) \psi_n =\lambda_n \psi_n$. This
equation's Green's function is $G(x,y) = \sum_{n} {\psi_n(x)
\psi_n^{\dagger}(y)\over \lambda_n}$.  If $\lambda$ is an eigenvalue
then $-\lambda$ is also an eigenvalue since $\lbrace\gamma_5 ,
i\raise.15ex\hbox{$/$}\kern-.53em\hbox{$D$} \rbrace =0$. This also
implies that the 2x2 matrix, $G$, is off-diagonal; so the trace of an
odd number of them vanishes.

The eigenfunctions of $i\raise.15ex\hbox{$/$}\kern-.53em\hbox{$D$}$
with zero eigenvalue are called zero modes.  They span the on-shell
part of the fermionic field and are dominant in the classical
limit. In general the gauge field may be written as $A_{\mu} =
-\varepsilon _{\mu \nu}\partial_{\nu}(\phi + f) + \partial_{\mu}
\rho$. The last term may be integrated over in the path integral with
no consequence. The function $f$ may be chosen to strictly carry any
singularity of the electric field, called a vortex configuration, and
$\phi$ is the quantum fluctuation around the vortex.  They appear in
the action as ${1\over4}F^2 = {1\over2}(\phi + f)\Box^2 (\phi + f)$.
{}From the index theorem \begin{displaymath} k={g\over 2\pi}\int
d^2x\hbox{ } E = {g\over 2\pi}\int d^2x \hbox{ } \varepsilon_{\mu\nu}
\partial_{\mu}A_{\nu} = {g\over 2\pi} \int d^2x\hbox{ } \Box
f,\end{displaymath} the background field, $f$, must satisfy
$f(x\rightarrow \infty) = {k\over g} \hbox{ ln} \vert x \vert$.
Therefore the strength of the vortex is related to $k$, the number of
right-handed minus the number of left-handed zero modes. The fact that
this quantity is not zero is due to the anomaly.

It can be further shown \cite{sw} that the zero modes of a vortex are
all of one chirality, hence $\vert k \vert$ is the total number of
zero modes.  This decomposition restricts integration of the gauge
field on the plane in the path integral to integration over $\phi$ and
a sum over all possible $k$.  The three gamma matrices $(\gamma_0 ,
\gamma_1 , \gamma_5 )$ can be chosen to be the Pauli matrices from
which the identity $\varepsilon_{\mu\nu} \gamma_{\nu} \gamma_{5}
=i\gamma_{\mu}$ follows.  Finally, solving
$i\raise.15ex\hbox{$/$}\kern-.53em\hbox{$D$} \chi =0$ for a given flux
on the plane gives \begin{eqnarray*} \chi^{(p)}_{+} (z,\overline z)
&=& N_{p,k}\, z^{p-1} e^{-g f(z,\overline{z})} \\ \chi^{(p)}_{-}
(z,\overline z) &=& N_{p,k}\, \overline{z}^{p-1} e^{g f(z,
\overline{z})}\end{eqnarray*} with $z=x^0 + i x^1$, $p=1,\ldots,|k|$,
and $N_{p,k}$ is the normalization.

\bigskip
{\bf 3. The Plane}
\bigskip

The $n$-point scalar correlator (hereafter called the $n$-$\sigma$
correlator) may be separated into sums of all possible combinations of
the chiral operators: \begin{equation}\overline{\Psi}(x)\Bigl({1\pm
\gamma_5\over2}\Bigr)\Psi(x)\equiv\overline{\Psi} P_{\pm}
\Psi(x)\end{equation} and evaluated using the ideas of Bardacki and
Crescimanno \cite{bc}.  A general combination of these operators
evaluated in a flux sector $k$ will be notated by: \begin{displaymath}
C^k_{e_1\ldots e_n}(x_1,\ldots,x_n)\equiv\Bigl\langle\prod_{i=1}^n
\overline{\Psi} P_{e_i}\Psi(x_i)\Bigr\rangle_k\end{displaymath} where
the $e_i$'s can be either plus or minus.  Since the Green's function
is off-diagonal, terms only contribute for $n-k$ even. The
transformation $\Psi(x)=e^{-g\gamma_5 \phi(x)}\chi(x)$ separates the
bosonic and fermionic integrals. Due to the chiral anomaly, this
transformation has a non-trivial Jacobian $e^{m^2\int
d^2x\phi\Box({1\over2}\phi+f)}$, which can be computed by building up
infinitesimal transformations on $\Psi$ (see Appendix A).

For the special case of $k=n$, only zero modes contribute to
$C^n_{+\ldots +}$. In fact, the fermionic integral will vanish unless
all the zero modes are taken into account here. A detailed calculation
given in Appendix B gives
\begin{displaymath}
C^n_{+\ldots+}(x_1,\ldots,x_n)=M_{(n)} e^{2g^2n
K_{xx}}e^{4g^2{\sum\atop i>j}K_{x_ix_j}} \, \left|\,
\begin{array}{*{4}{c}} 1&z_1&\ldots&z_1^{n-1} \\
                 1&z_2&\ldots&z_2^{n-1} \\ \,\vdots\hfill&&&\,\vdots
                 \\ 1&z_n&\ldots&z_n^{n-1}\end{array} \right|^2=
\end{displaymath}
\begin{displaymath}
=M_{(n)} e^{2g^2n K_{xx}}e^{4g^2{\sum\atop i>j}K_{x_ix_j}}
\prod_{i>j=1}^{n} |z_i-z_j|^2.
\end{displaymath}
The last equality can be proven by setting various $z_i$'s and $z_j$'s
equal in the determinant and noting this is a zero since two lines are
then equal.  Here $K_{xy}$ is the bosonic Green's function. It can be
calculated by introduction of an infrared regulator, $\mu$, which will
be taken to zero at the end of the calculation.
\begin{displaymath}
K_{xy}={1\over\Box^2-m^2\Box}\delta^2(x-y)={1\over
m^2}\Bigl({1\over\Box-m^2}-{1\over\Box}\Bigr)\delta^2(x-y)=\end{displaymath}
\begin{displaymath} =\lim_{\mu\to0}{1\over
m^2}\Bigl({1\over\Box-m^2}-{1\over\Box-\mu^2}\Bigr)
\delta^2(x-y)=-{1\over2g^2}\bigl[K_0(m|x-y|)+\ln{\mu e^\gamma\over2}|x-y|
\bigr]
\end{displaymath}
\begin{displaymath}
K_{xx}={1\over2g^2}\ln{m\over\mu}
\end{displaymath}
Since $\mu$ does not depend on $k$, it is the same parameter for all
Green's functions. Substituting in for all terms, the $|z_i-z_j|^2$
terms from the zero mode determinant cancel with the terms from the
Green's functions giving \begin{equation}
C^k_{+\ldots+}(x_1,\ldots,x_k)=M_{(k)}{m^k \over \mu^{k^2}}
\Bigl({2\over e^\gamma}\Bigr)^{k(k-1)} e^{-2{\sum\atop{i>j}}
K_0^{x_ix_j}}=C^{-k}_{-\ldots-}(x_1,\ldots,x_n).\end{equation} where
$K_0^{xy}\equiv K_0(m|x-y|)$ and $M_{(k)}\equiv N_{1,k}^{\ 2}\ldots
N_{k,k}^{\ 2}$.

By taking $|x_i-x_j|\to\infty$ for all $i$ and $j$, the cluster
approximation reduces this correlator to a form only depending on the
condensate \cite{sw}, $\langle\overline{\Psi}\Psi\rangle=-me^\gamma
/2\pi$

\begin{displaymath}
C^k_{+\ldots+}(x_1,\ldots,x_n)\,\rightarrow\,
\Bigl({\langle\overline{\Psi}\Psi\rangle\over2}\Bigr)^k=M_{(k)}
{m^k\over\mu^{k^2}}
\Bigl({2\over e^\gamma}\Bigr)^{k(k-1)},
\end{displaymath}
and so a general form for $M_{(k)}$ may be found.
\begin{equation}
M_{(k)}=\Bigl({-1\over 2\pi}\Bigr)^k\Bigl({\mu e^\gamma\over2}
\Bigr)^{k^2}\label{cluster}
\end{equation}
On the plane, the $p=k$ zero mode is non-normalizable.  This takes the
form of an infrared (large distance) divergence. However, this is
exactly canceled by the infrared (small momentum) divergence of
$(\Box-\mu^2)^{-1}$ as the regulator $\mu\rightarrow 0$ as seen in
equation \ref{cluster}.

Calculation of non-minimal correlation functions requires knowledge of
the fermionic Green's function \cite{bc} of
$i\raise.15ex\hbox{$/$}\kern-.53em\hbox{$\tilde
D$}=i(\raise.15ex\hbox{$/$}\kern-.57em\hbox{$\partial$} - g\gamma_5
\raise.15ex\hbox{$/$}\kern-.57em\hbox{$\partial$} f)$ in arbitrary
flux $k$.  For convenience, we choose $k>0$.
\begin{displaymath}
i\raise.15ex\hbox{$/$}\kern-.53em\hbox{$\tilde D$}
G(x,y)=\delta^2(x-y)-P(x,y)
\end{displaymath}
\begin{displaymath}
\hfill \Rightarrow  \left\{
\begin{array}{c@{\:=\:}c}
2i[\partial_z - g(\partial_z f)] G_{-+}(x,y)&\delta^2(x-y)-P(x,y)\\
2i[\partial_{\overline z} + g(\partial_{\overline z}f)]G_{+-}(x,y)&
\delta^2(x-y)\end{array}\right.
\end{displaymath}
\begin{eqnarray*}
G_{-+}(x,x') =\bigl\langle 0\bigr|
\Psi_{L}(x)\overline{\Psi}_{R}(x')\bigl| 0 \bigr\rangle =
{e^{g[f(z,{\overline z})-f(z',{\overline z}')]}\over 2\pi i
({\overline z} - {\overline z}')};&& G_{--}=0 \\ G_{+-}(x,x')
=\bigl\langle 0\bigr| \Psi_{R}(x)\overline{\Psi}_{L}(x')\bigl| 0
\bigr\rangle = {e^{g[f(z',{\overline z}')-f(z,{\overline z})]}\over
2\pi i (z - z')};&& G_{++}=0.
\end{eqnarray*}
Here $P(x,y)$ is the projection operator onto the space of zero
modes. The zero modes are all right handed and therefore $P(x,y)$
appears in only the $G_{-+}$ equation. For $k<0$ the above equations
would be exactly the same except that $P(x,y)$ would only appear in
the $G_{+-}$ equation. In fact, $P(x,y)$ may be ignored in the
calculation of correlation functions since the product of zero modes
and contractions are antisymmetrized \cite{bc} as expounded upon in
Appendix C.  In general, an analogue of Wick's theorem may be used to
write out all contractions of $\Psi_{R}$'s and $\Psi_{L}$'s
possible. The chiral neutral terms (like
$\Psi_{R}\overline{\Psi}_{L}$) can be replaced by the corresponding
Green's function above, and the purely chiral term with all the
$\Psi_{\bigl\{{R\atop L}\bigr\}}$'s (for $k=\pm|k|$) can be replaced
by the zero modes.

Since $G_{+-}(x,y)=G^{*}_{-+}(y,x)$, the general non-minimal
correlator may be written as (for $r-s=k>0$ and $r+s=n$):
\begin{displaymath}
C^k_{(+\ldots+)(-\ldots-)}(x_1,\ldots,x_r;x_{r+1},\ldots,x_{r+s})=
e^{2g^2n K_{xx}} e^{4g^2{\sum\atop i>j}e_ie_jK_{x_ix_j}}
e^{2g{{n\atop\sum}\atop i=1}e_if(x_i)}\times
\end{displaymath}
\begin{equation}
\times \, \left|\,
\begin{array}{*{6}{c}} \chi_+^1(z_1,{\overline z}_1)&\ldots&
\chi_+^k(z_1,{\overline
z}_1)&G_{+-}(x_1,x_{r+1})&\ldots&G_{+-}(x_1,x_{r+s}) \\
\,\vdots\hfill&&&&&\,\vdots \\ \chi_+^1(z_r,{\overline
z}_r)&\ldots&\chi_+^k(z_r,{\overline
z}_r)&G_{+-}(x_r,x_{r+1})&\ldots&G_{+-}(x_r,x_{r+s})\end{array}\right|^2
\end{equation}

with $e_i=\,+$ for $1\le i\le r$ and $e_i=\,-$ for $r+1\le i\le
r+s$. With this configuration,
\begin{displaymath}
\sum_{i>j}e_ie_j={(r-s)^2-(r+s)\over2}={k^2-n\over2}.
\end{displaymath}
Writing out explicitly all the terms, equation (4) becomes
\begin{displaymath}
M_{(k)}\Bigl({m\over\mu}\Bigr)^n\Bigl({2\over\mu
e^\gamma}\Bigr)^{k^2-n} {1\over(2\pi)^{2s}} e^{-2{\sum\atop
i>j}e_ie_jK_0^{x_ix_j}} e^{2g{{n\atop\sum}\atop i=1}e_if(x_i)}
\prod_{i>j}|z_i-z_j|^{-2e_ie_j}\times
\end{displaymath}
\begin{displaymath}
\hfill\times
\, \left|\,\begin{array}{*{7}{c}} e^{-gf(x_1)}&z_1e^{-gf(x_1)}&\ldots
&z_1^{\ k-1}e^{-gf(x_1)} &{e^{g[f(x_{r+1})-f(x_1)]}\over
z_1-z_{r+1}}&\ldots&{e^{g[f(x_{r+s})-f(x_1)]} \over z_1-z_{r+s}}\\
\,\vdots\hfill&&&&&&\,\vdots\\
e^{-gf(x_r)}&z_re^{-gf(x_r)}&\ldots&z_r^{\ k-1}e^{-gf(x_r)}&
{e^{g[f(x_{r+1})-f(x_r)]}\over
z_r-z_{r+1}}&\ldots&{e^{g[f(x_{r+s})-f(x_r)]} \over
z_r-z_{r+s}}\end{array}\right|^2
\end{displaymath}
It is now easy to see how the $f$ dependence from the determinant
cancels the extra term from the bosonic integration. Also, as in the
minimal correlator above, the $|z_i-z_j|^2$ terms cancel as
well. Since the only allowed $k$ are such that $(-1)^k=(-1)^n$, the
remaining terms may be written as:
\begin{displaymath}
C^k_{(+\ldots+)(-\ldots-)}(x_1,\ldots,x_n)=\Biggl(-{me^\gamma\over4\pi}
\Biggr)^n e^{-2{\sum\atop i>j}e_ie_jK_0^{x_ix_j}}.
\end{displaymath}

This can be inserted into the correlators to give
\begin{eqnarray}
\Bigl\langle \overline{\Psi}\Psi(x) \overline{\Psi}\Psi(y) \Bigr\rangle
&=&\langle\overline{\Psi}\Psi \rangle^2 \cosh \bigl( 2
K_0^{xy}\bigr)\nonumber\\ \Bigl\langle \overline{\Psi}\Psi(x)
\overline{\Psi}\Psi(y) \overline{\Psi}\Psi(z) \Bigr\rangle &=&
\langle\overline{\Psi}\Psi\rangle^3 {1\over4}\bigl(e^{2(K_0^{xz}+
K_0^{yz}-K_0^{xy})}+e^{2(K_0^{yz}+K_0^{xy}-K_0^{xz})}+\nonumber\\
\hfill &\quad&
+e^{2(K_0^{xy}+K_0^{xz}-K_0^{yz})}+e^{-2(K_0^{xy}+K_0^{xz}+K_0^{yz})}
\bigr)\nonumber\\ \hfill &=&\langle\overline{\Psi}\Psi\rangle^3
[\cosh(2K_0^{xy})\cosh(2K_0^{xz})\cosh(2K_0^{yz})+\nonumber\\ &\quad&
-\sinh(2K_0^{xy})\sinh(2K_0^{xz})\sinh(2K_0^{yz})]\nonumber\\
\Bigl\langle \prod_{i=1}^n \overline{\Psi}\Psi(x_i)\Bigr\rangle &=&
\Bigl({\langle\overline{\Psi}\Psi\rangle\over2}\Bigr)^n\sum_{\{ e_i\}
\in \{\pm\} } e^{-2{\sum\atop
i>j}e_ie_jK_0^{x_ix_j}}.\label{final}\end{eqnarray} The last equation,
(\ref{final}) is the general $n$-$\sigma$ correlator. The sum for the
sets of $e_i$'s runs over all possible combinations of pluses and
minuses.  It is interesting to note that this expression can be viewed
as the partition function of a two dimensional system of $n$ fermions
at fixed points with charges $\pm1$ and potential $K_0^{xy}$ which at
small distances reduces to the potential studied by Kosterlitz and
Thouless \cite{kt}. This observation was originally due to Samuel
\cite{sam}.

\bigskip
{\bf 4. The Torus}
\bigskip

We would like to study the Schwinger model at finite
temperature. Placing the model on a torus \cite{sw} with sides
$[0,L]\times[0,\beta]$ gives the appropriate geometry for such
studies.  Finally, we shall impose the thermodynamic limit by taking
$L$ to infinity and obtain the finite temperature result. Taking
$\beta$ to infinity should reproduce the results obtained on the plane
for zero temperature.

The zero modes are more complicated here than on the plane (Jacobi
theta functions) yet are normalizable in the finite volume.
\begin{displaymath}
\varphi_{p}(x)=\Bigl({2|k|\over\beta^2 V}\Bigr)^{1 \over4} U(x)\,
e^{\mp g[\phi (x) + f(x)]} \, \Theta\Bigl[{(p-{1\over2}-h_0)/k\atop
h_1}\Bigr](kz,i|k|\tau);
\end{displaymath}
\begin{displaymath}
\hbox{with}\quad f(x)={\pi k\over
V}(x^1)^2,\quad U(x)=e^{2\pi i[h_0 x^0/\beta+h_1 x^1/L]},
\end{displaymath}
$z=(x^0+ix^1)/\beta$, and $\tau=L/\beta$ similar to the notation of
Sachs and Wipf \cite{sw}.  The singularity that $\Box f$ had on the
plane is spread out over the torus so as not to ruin the
periodicity. The $A$ integration may be replaced by a sum over $k$ and
integration over $\phi$ as in Appendix B for the plane. In addition,
$h_0$ and $h_1$ are the $2\pi$ periodic degrees of freedom gained by
changing the topology of the space and they must also be integrated
over:
\begin{displaymath}
\int DA_{\mu}\ldots=\sum_k\langle\ldots\rangle_A=\sum_k\int
d^2h D\phi\ldots e^{-{1\over4}\int F^2}=
\end{displaymath}
\begin{displaymath}
=\sum_k\int d^2h D\phi\ldots e^{-{1\over2}\int\phi\Box^2\phi} e^{-2\pi
k^2/m^2V}.
\end{displaymath}
The $h$'s are defined modulo $1$ corresponding to a full revolution
around the two periodicities of the torus.

The 3-$\sigma$ correlator may be broken up into four distinct terms
considering the equivalent contributions from the positive and
negative $k$ sectors.
\begin{displaymath}
\Bigl\langle\overline{\Psi}\Psi(x)\overline{\Psi}\Psi(y)
\overline{\Psi}\Psi(z)\Bigr\rangle=
2[C^3_{+++}(x,y,z)+C^1_{++-}(x,y,z)+C^1_{++-}(y,z,x)+C^1_{++-}(z,x,y)].
\end{displaymath}
The general form of these terms may be written conveniently as (for
$r-s=k>0$):
\begin{displaymath}
C^k_{(+\ldots+)(-\ldots-)}(x_1,\ldots,x_r;x_{r+1},\ldots,x_{r+s})=
\end{displaymath}
\begin{displaymath}
\hfill=Z^{-1}\Bigl\langle \hbox{det}{'}
(i\raise.15ex\hbox{$/$}\kern-.53em\hbox{$D$}_k)
\, \left|\,\begin{array}{*{6}{c}}
\varphi_1(x_1)&\ldots&\varphi_k(x_1)&S_{+-}(x_1,x_{r+1})
&\ldots &S_{+-}(x_1,x_{r+s})\\ \,\vdots\hfill&&&&&\,\vdots\\
\varphi_1(x_r)&\ldots&\varphi_k(x_r)&S_{+-}(x_r,x_{r+1})&\ldots&S_{+-}
(x_r,x_{r+s})\end{array}\right|^2\Bigr\rangle_A.
\end{displaymath}

The finite temperature Green's function in the presence of flux $k$
satisfies the equation:
\begin{displaymath}
i\raise.15ex\hbox{$/$}\kern-.53em\hbox{$D$}
S(x,y)=\delta^2(x,y)-P(x,y).
\end{displaymath}
Appendix C shows that the solution can be written as just the $k=0$
Green's function \cite{zahed} with an $h$-independent term multiplying
it to ensure correct boundary conditions.
\begin{displaymath}
S_{+-}(x,y)=S_{-+}^{*}(y,x)=
\end{displaymath}
\begin{displaymath}
={1\over 2\pi i\beta} e^{g(\phi(y)-\phi(x)+f(y)-f(x))} U(x)
U^{\dagger}(y)e^{-2\pi i h_0(z_1-z_2)} \vartheta_1 '(0)T_k(z_1,z_2)
\end{displaymath}
\begin{displaymath}
\hbox{where}\quad T_k(z_1,z_2)=
{\vartheta_3(kz_1|ik\tau)\vartheta_4(z_1-z_2+H) \over
\vartheta_3(kz_2|ik\tau)\vartheta_4(H)\vartheta_1(z_1-z_2)}\quad
\hbox{and}\quad H=h_1-i\tau h_0.
\end{displaymath}
These theta functions are related to those from Sachs and Wipf by:
\begin{displaymath}
\Theta\Bigl[{a\atop b}\Bigr](z,i\tau)=e^{2\pi i a(z+b+{i\over2}\tau a)}
\vartheta_3(\pi(z+b+i\tau a)|i\tau)
\end{displaymath}
and the $\pi$ in the argument and the dependence on $\tau$ has been
suppressed for ease in reading.

Continuing the calculation of the 3-$\sigma$ correlator, it may be
simplified after the $\phi$ integration to
\begin{displaymath}
C^1_{++-}(x,y,z)={\sqrt{2\tau}|\eta(i\tau)|^8\over\beta^3}\,
e^{6g^2K_{xx}+4g^2(K_{xy}-K_{xz}-K_{yz})} e^{-2\pi/m^2V}\times
\end{displaymath}
\begin{displaymath}
\hfill\times\int d^2h e^{-2\pi \tau h_0^{\ 2}+4\pi
h_0(x^1+y^1-z^1)/\beta} \, \left|\,\begin{array}{*{2}{c}}
\vartheta_2(z_1+H)&T_1(z_1,z_3)\\
\vartheta_2(z_2+H)&T_1(z_2,z_3)\end{array}\right|^2.
\end{displaymath}
Here $K_{xy}$ is the bosonic Green's function as before on the plane.
Its explicit form on the torus is (see Appendix D)
\begin{displaymath}
4g^2 K_{xy}=4\pi\Delta_m(x-y)+\ln
|{\eta(i\tau)\over\vartheta_1(z_1-z_2)}|^2 + {2\pi\over V} (x^1-y^1)^2
+ {4\pi\over m^2 V}.
\end{displaymath}
After also identifying the finite temperature condensate \cite{sw}:
\begin{displaymath}
\langle\overline{\Psi}\Psi\rangle_{\beta}=-{2|\eta(i\tau)|^2\over\beta}
e^{-2\pi/ m^2V} e^{2g^2 K_{xx}},
\end{displaymath}
the final form for finite volume is
\begin{equation}
C^1_{++-}(x,y,z)=\biggl({\bigl\langle\overline{\Psi}\Psi\bigr
\rangle_{\beta}\over2}
\biggr)^3 \, e^{4\pi[\Delta_m(x-y)-\Delta_m(x-z)-\Delta_m(y-z)]}
F_{++-}(x,y,z)
\label{finite}
\end{equation}
with
\begin{displaymath}
F_{++-}(x,y,z)=
{\sqrt{2\tau}\over|\vartheta_1(z_1-z_2)\vartheta_3(z_3)|^2} \int
{d^2h\over|\vartheta_4(H)|^2} e^{-2\pi(x^1+y^1-z^1-Lh_0)^2/V} \times
\end{displaymath}
\begin{displaymath}
\hfill\times \, \left|\,\begin{array}{*{2}{c}}
\vartheta_2(z_1+H)&\vartheta_3(z_1)\vartheta_4(z_1-z_3+H)
\vartheta_1(z_2-z_3)\\ &\hfill\\
\vartheta_2(z_2+H)&\vartheta_3(z_2)\vartheta_4(z_2-z_3+H)\vartheta_1(z_1-z_3)
\end{array}\right|^2.
\end{displaymath}
The full massive bosonic Green's function, $2\pi\Delta_m(x-y)$ may be
rewritten in the limit of $L\to\infty$ giving :
\begin{displaymath}
-2\pi\Delta_m(x)=\sum_{k=-\infty}^{\infty} K_0(m\sqrt{(x^0-\beta k)^2
+ (x^1)^2})
\end{displaymath}
which shows the zero temperature contribution is just $K_0(m|x|)$ as
obtained on the plane. There is also an extra topological factor,
$F_{++-}$, which is not able to be exactly integrated, but as
$L\to\infty$:
\begin{eqnarray*}\vartheta_1(z)\to 2 e^{-{\pi\tau\over4}}\sin
\pi z\quad &&\vartheta_2
(z+H)\to e^{-\pi\tau({1\over4}-h_0)} e^{\pi i(z+h_1)}\\
\vartheta_3(z)\to 1 &&\vartheta_4(z+H)\to 1-e^{\pi\tau(2h_0-1)}
e^{2\pi i(z+h_1)}.\end{eqnarray*} and $F_{++-}(x,y,z)\to 1$.  Taking
$\beta\to\infty$ gives the same result as on the plane, thereby
confirming our results.

Considering ways in which the general result could depend on $\beta$
such that the zero temperature limit is equation (\ref{final}) and the
$3$-$\sigma$ correlator is equation (\ref{finite}) compel us to one
conclusion. This is that the finite temperature result can be obtained
from the zero temperature result by simply replacing the zero
temperature condensate and Green's function with their finite
temperature counterparts. If this is indeed the case, the $n$-$\sigma$
finite temperature correlator can then be written as
\begin{displaymath}
\Bigl\langle\prod_{i=1}^n\overline{\Psi}\Psi(x_i)\Bigr\rangle= \left(
{\bigl\langle\overline{\Psi} \Psi\bigr\rangle_{\beta}\over2}\right)^n
\sum_{\{e_i\}\in\{\pm\}} e^{4\pi {\sum\atop i>j} e_i
e_j\Delta_m(x_i-x_j)}.
\end{displaymath}
In the bosonized form of the Schwinger model \cite{coleman} the
structure of this result is expected \cite{smilga2}. It is reassuring
that the exact finite temperature calculation done here confirms this,
giving us some confidence in the generalization procedure at finite
temperature and infinite volume. The generalization to finite
temperature and finite volume is less straightforward.

\bigskip
{\bf 5. Conclusion}
\bigskip

Two important ways to calculate the $n$-$\sigma$ correlation functions
in the Schwinger model were looked at specifically to find a general
form for the $n$-point scalar correlators. The plane is easier to work
on due to the absence of the cumbersome theta functions, but the
validity of results from the plane have been questioned in the
past. Therefore the result was also derived on the torus and shown to
be equivalent to the plane in the large volume limit.

In fact, both methods are very similar. The large distance divergence
in the zero modes are canceled by the small momentum divergence in
the bosonic Green's function. The final result for the finite
temperature $n$-$\sigma$ correlator is identical to the zero
temperature result if the zero temperature condensate and Green's
function are replaced by their finite temperature counterparts. There
is no other spurious temperature dependence as could be expected due
to the differing topologies in which the two problems were solved.

The results presented here are directly amenable to the spectral
analysis discussed by Fayyazuddin $et$ $al.$ \cite{zahed}. They show
that in the Schwinger model, the four-, six-, ..., $n$-point scalar
correlation functions have a screening length that is determined by
the meson mass $m=e/\sqrt{\pi}$ and independent of the
temperature. This is not the case of the two-point gauge invariant
fermion correlator, where the screening mass was found to asymptote
$\pi T/2$ due to the emergence of two-meron configurations at high
temperature \cite{jvz}.  While the present calculations are schematic
and far from the real world, they still provide nontrivial insights
into nonperturbative physics at finite temperature. They are certainly
of some interest for comparison with finite temperature lattice
simulations of the model \cite{latt}.

\vglue 0.6cm
{\bf \noindent  Acknowledgements \hfil}
\vglue 0.4cm
This work was supported in part by the US DOE grant DE-FG02-88ER40388.

\bigskip
{\bf Appendix A}
\bigskip

The non-invariance of the fermionic measure under the chiral
transformation $\Psi(x)= e^{-g\gamma_5 \phi(x)} \chi(x)$ leads to a
non-trivial Jacobian.  First noticed by Fujikawa in four
dim\-en\-sions \cite{fuji}, we give a derivation in the two
dimensional case because there are certain subtleties and we have not
seen an explicit calculation in the literature.

Expanding $\Psi(x)$ and $\chi(x)$ in orthonormal eigenmodes of the
Dirac operator, $\varphi_n(x)$, the above transformation for an
infinitesimal change, $\delta\phi(x)$, in the quantum fluctuation field
gives
\begin{displaymath} D\overline{\Psi} D\Psi = \exp\Bigl[-\int d^2x\,
\delta\phi(x)\,
\sum_n\varphi_n^{\dagger}(x)\gamma_5\varphi_n(x)\Bigr]
D\overline{\chi} D\chi=\end{displaymath}
\begin{displaymath}=e^{m^2\int d^2x\,\delta\phi\,
\varepsilon_{\mu\nu}\partial_{\mu}A_{\nu}}
D\overline{\chi} D\chi\end{displaymath} for the measure. However,
$A_\nu=-\varepsilon_{\nu\rho}\partial_\rho (\delta\phi + f)$ and so
\begin{displaymath}\delta\phi\varepsilon_{\mu\nu}\partial_\mu A_\nu
=\delta\phi\Box(\delta\phi
+f).\end{displaymath} Integrating the infinitesimal transformations,
the additional term in the Lagrangian for finite $\phi$ is $m^2\int
d^2x\phi\Box({1\over2}\phi+f)$.

\bigskip
\newpage
{\bf Appendix B}
\bigskip

The $n$-$\sigma$ correlator may be broken into $s_{+}$ and $s_{-}$
parts and simplified by the chiral transformation discussed in
Appendix A to separate the fermionic and bosonic integrations.
\begin{displaymath} C^k_{e_1\ldots e_n}(x_1,\ldots,x_n) =
Z^{-1} \int D\overline{\chi} D\chi DA \prod_{i=1}^{n} \overline{\chi}
P_{e_i} \chi (x_i)\,e^{-2g\sum e_i\phi (x_i)}\,
e^{-S_{eff}}\end{displaymath}
\begin{displaymath}\hfill=Z^{-1}\sum_{k}\int D\overline{\chi}
D\chi D\phi \prod_{i=1}^{n} \overline{\chi}
P_{e_i}\chi (x_i)\, e^{-\int d^2x(-\overline{\chi}
i{\raise.15ex\hbox{$/$}\kern-.53em\hbox{$\tilde D$}}
\chi+J\phi+{1\over2}f\Box^2 f)}\, e^{-{1\over2}\int
d^2x\phi(\Box^2-m^2\Box)\phi}\end{displaymath}
\begin{displaymath}\hbox{where}\quad S_{eff}=\int d^2x
\Bigl[-\overline{\chi}
i {\raise.15ex\hbox{$/$}\kern-.53em\hbox{$\tilde D$}} \chi+
{1\over2}(\phi+f)(\Box^2-m^2\Box)(\phi+f) + {m^2\over2}f\Box
f\Bigr]\end{displaymath}
\begin{displaymath} J=(\Box^2-m^2\Box)f+2g\sum_{i=1}^n
e_i\delta^2(x-x_i)\qquad
\hbox{and}\qquad{\raise.15ex\hbox{$/$}\kern-.53em\hbox{$\tilde D$}}
=\raise.15ex\hbox{$/$}\kern-.57em\hbox{$\partial$}
-g\gamma_5\raise.15ex\hbox{$/$}\kern-.57em\hbox{$\partial$} f.
\end{displaymath}
Now completing the square in the bosonic integral (with
$B=\Box^2-m^2\Box$):
\begin{displaymath}{1\over2}\phi B\phi + J\phi={1\over2}\bigl
(\phi+{J\over B}\bigr)B\bigl(\phi+
{J\over B}\bigr)-{1\over2}JB^{-1}J.\end{displaymath} Therefore the
bosonic Gaussian cancels the same factor in the denominator.  Further
calculation yields:
\begin{displaymath}{1\over2}\int d^2xJB^{-1}J ={1\over2}\int d^2x fBf+
2g\sum_{i=1}^{n} e_i f(x_i) + 2g^2\sum_{i,j}^{n} e_i e_j K_{x_i
x_j}\end{displaymath}
\begin{displaymath}={1\over2} \int d^2x f(\Box^2-m^2\Box)f +
 2g\sum_{i=1}^{n} e_i f(x_i) + 2g^2 n
K_{xx} + 4g^2\sum_{i>j}^{n} e_i e_j K_{x_ix_j}.\end{displaymath}

Antisymmetrizing the fermionic functions:
\begin{displaymath}\sum_{\sigma} (-1)^{\sigma}\prod_{p=1}^k
\chi_{R}^{(p)}(x_{\sigma(p)})=
M_{(k)}^{1/2} e^{-g\sum f(x_i)}\, \left|\,\begin{array}{*{4}{c}}
1&z_1&\ldots&z_1^{k-1}\\ 1&z_2&\ldots&z_2^{k-1}\\
\,\vdots\hfill&&&\,\vdots\\
1&z_k&\ldots&z_k^{k-1}\end{array}\right|=\end{displaymath}
\begin{displaymath}=M_{(k)}^{1/2} e^{-g\sum f(x_i)}
\prod_{i>j=1}^{k} (z_i-z_j).\end{displaymath}
The non-zero mode fermionic integration leads to a factor:
\begin{displaymath}
{{\det'(i{\raise.15ex\hbox{$/$}\kern-.53em\hbox{$\tilde D$}}_k)}
\over{\det(i\raise.15ex\hbox{$/$}\kern-.57em\hbox{$\partial$})}}
=\exp\Bigl[{1\over2}\hbox{Tr}\ln
\Bigl({-{\raise.15ex\hbox{$/$}\kern-.53em\hbox{$\tilde D$}}^2\over
-\raise.15ex\hbox{$/$}\kern-.57em\hbox{$\partial$}^2}\Bigr)\Bigr]
=e^{{m^2\over2}\int d^2x f\Box f}\end{displaymath}
since
\begin{displaymath}\hbox{Tr}\ln\Bigl(
{{-{\raise.15ex\hbox{$/$}\kern-.53em\hbox{$\tilde D$}}^2}
\over{-\raise.15ex\hbox{$/$}\kern-.57em\hbox{$\partial$}^2}}\Bigr)
=-\hbox{Tr}\int_0^1
d\alpha {d\over d\alpha} \int_0^{\infty}{ds\over s}
e^{s{\raise.15ex\hbox{$/$}\kern-.53em\hbox{$\tilde D$}}_{\alpha}^2}
=m^2\int d^2x f\Box f\end{displaymath} where
${\raise.15ex\hbox{$/$}\kern-.53em\hbox{$\tilde
D$}}_{\alpha}=e^{\alpha g \gamma_5
f}\,\raise.15ex\hbox{$/$}\kern-.57em\hbox{$\partial$}\, e^{\alpha g
\gamma_5 f}$.

Gathering all terms together, the zero mode saturated correlator
becomes
\begin{eqnarray*}\Bigl\langle\prod_{i=1}^k s_{+} (x_i)\Bigr\rangle_k &=&
M_{(k)}\prod_{i>j} |x_i-x_j|^2\, e^{2 g^2 k K_{xx}}\,
e^{4g^2{\sum\atop{i>j}} K_{x_i x_j}}\\ &=& M_{(k)} {m^k\over\mu^{k^2}}
\Bigl({2\over e^{\gamma}}\Bigr)^{k(k-1)} e^{-2{\sum\atop{i>j}}
K_0(m|x_i-x_j|)}.\end{eqnarray*} where $|z_i-z_j|=|x_i-x_j|$ was used.

\bigskip
{\bf Appendix C}
\bigskip

We derive the fermionic Green's function in the presence of a
non-trivial flux $k$ on the torus. This is required for the
non-minimal correlation functions at finite temperature. Although
other authors have tackled this problem \cite{stam},
a more explicit form is
needed for practical calculations. The Green's function is defined by:
\begin{displaymath} i\raise.15ex\hbox{$/$}\kern-.53em\hbox{$D$}
S(x,y)=\delta^2(x-y)-P(x,y)\end{displaymath}
where $P(x,y)$ is the projection operator onto the space of zero
modes. For convenience, $k$ will be taken positive.

Using the trivialization of the $U(1)$ bundle on the torus as in Sachs
and Wipf \cite{sw}, the gauge potential is explicitly:
\begin{eqnarray*} A_0&=&-{\Phi\over V}x^1+{2\pi\over\beta}h_0-
\partial_1\phi\\
A_1&=&{2\pi\over L}h_1+\partial_0\phi.\end{eqnarray*}
Noticing that:
\begin{equation} i\raise.15ex\hbox{$/$}\kern-.53em\hbox{$D$}=
e^{g\gamma_5[\phi(x)+f(x)]}U(x)i\raise.15ex\hbox{$/$}\kern-.57em
\hbox{$\partial$} U^{\dagger}(x)
e^{g\gamma_5[\phi(x) +f(x)]},\label{ds}\end{equation} we can write
\begin{equation} S(x,y)=e^{-g\gamma_5[\phi(x)+f(x)]} U(x) g(x,y)
U^{\dagger}(y)
e^{-g\gamma_5 [\phi(y)+f(y)]}.\label{s}\end{equation} This reduces our
problem to solving the equation:
\begin{displaymath} i\raise.15ex\hbox{$/$}\kern-.57em\hbox{$\partial$}
g(x,y)=\delta^2(x-y)-{\tilde P}(x,y)\end{displaymath}
\begin{displaymath}\hbox{where}\quad {\tilde P}(x,y)
=e^{g\gamma_5[\phi(x)+f(x)]}U^{\dagger}(x)
P(x,y) U(y) e^{g\gamma_5[\phi(y)+f(y)]}.\end{displaymath}

Writing the components of this matrix equation out and letting
$z=(x^0+ix^1)/\beta$ and $w=(y^0+iy^1)/\beta$,
\begin{eqnarray*} -\beta \partial_{z} g_{-+}(z-w)&=&\delta^2(z-w)-
{\tilde P}(z,w)\\ -\beta\partial_{\overline z}
g_{+-}(z-w)&=&\delta^2(z-w).\end{eqnarray*} Following the ideas of
\cite{bc}, we realize that $g_{+-}$ and $g_{-+}$ both appear as
contractions of left- and right-handed fermionic fields in
calculations of the non-minimal correlation functions. Writing
\begin{displaymath} g_{-+}(z,w)={\tilde g}_{-+}(z,w)-\sum_{n=1}^k
\chi_n(z) \eta^{*}_n(w)\end{displaymath}
\begin{displaymath}\hbox{with}\qquad \sum_{n=1}^k
\eta_n(z)\eta^{*}_n(w)
={\tilde P}(z,w),\end{displaymath}
\begin{displaymath} -\beta\partial_{z} \chi_n(z)=\eta_n(z),
\quad\hbox{and}\quad
-\beta\partial_{z} {\tilde
g}_{-+}(z,w)=\delta^2(z-w).\end{displaymath} The summation above is a
summation of the zero modes. In the calculation of non-minimal
correlation functions, the product of zero modes and contractions
needs to be antisymmetrized. As shown in section {\bf 4}, the Green's
function appears along with all of the zero modes in a
determinant. For this reason, the extra summation above will not
contribute. Therefore only ${\tilde g}_{+-}$ and ${\tilde g}_{-+}$
need to be calculated.

In the presence of a non-zero flux, fields at $(x^0,x^1)$ and
$(x^0,x^1+L)$ are related by the gauge transformation:
\begin{displaymath} A_{\mu}(x^0,x^1+L)-A_{\mu}(x^0,x^1)
=\partial_{\mu}\alpha,\qquad\Psi(x^0,
x^1+L)=e^{i\alpha}\Psi(x^0,x^1)\end{displaymath}
\begin{equation}\hbox{where}\qquad \alpha
=-{2\pi k\over\beta}x^0.\label{a}\end{equation}
{}From equations \ref{s} and \ref{a}, the boundary conditions on
${\tilde g}$ can be found to be:
\begin{displaymath}{\tilde g}_{+-}(z+1,w)=
-e^{-2\pi i h_0} {\tilde g}_{+-}(z,w)\qquad
{\tilde g}_{+-}(z,w+1)=-e^{2\pi i h_0} {\tilde
g}_{+-}(z,w)\end{displaymath}
\begin{equation}{\tilde g}_{+-}(z+i\tau,w)=
e^{\pi k\tau} e^{-2\pi i (kz+h_1)}
{\tilde g}_{+-}(z,w) \label{bound}\end{equation}
\begin{displaymath}{\tilde g}_{+-}(z,w+i\tau)=
e^{-\pi k\tau} e^{2\pi i (kw+h_1)}
{\tilde g}_{+-}(z,w).\end{displaymath}

The general form of ${\tilde g}_{+-}$ is
\begin{displaymath}{\tilde g}_{+-}(z,w)=
{1\over 2\pi i\beta} {\vartheta_1 '(0)\over\vartheta_1
(z-w)} f(z,w)\end{displaymath} and $f(z,w)$ must be chosen to complete
the boundary conditions in equation \ref{bound}, have no poles for
$z=w$, and be $1$ at $z=w$.  A solution that satisfies these
conditions is
\begin{displaymath} f(z,w)=
{\vartheta_4(z-w+H)\over\vartheta_4(H)}{\vartheta_3(kz|ik\tau)\over
\vartheta_3(kw|ik\tau)}e^{-2\pi i h_0(z-w)}.\end{displaymath}
A similar calculation can be done for ${\tilde g}_{-+}$ giving
finally:
\begin{displaymath}{\tilde g}_{+-}(z,w)=
{1\over 2\pi i \beta}{\vartheta_1 '(0)\vartheta_4(z-w+H)
\vartheta_3(kz|ik\tau)\over\vartheta_1(z-w)\vartheta_4(H)
\vartheta_3(kw|ik\tau)}e^{-2\pi i h_0(z-w)}\end{displaymath}
\begin{displaymath}{\tilde g}_{-+}(z,w)=
{1\over 2\pi i \beta}{\vartheta_1 '(0)\vartheta_4(
{\overline z}-{\overline w}-{\overline H}) \vartheta_3(k{\overline
w}|ik\tau)\over\vartheta_1({\overline z}-{\overline w})
\vartheta_4({\overline H}) \vartheta_3(k{\overline z}|ik\tau)}e^{-2\pi
i h_0({\overline z}-{\overline w} )}.\end{displaymath} Inserting this
in equation \ref{s}, we obtain the fermionic Green's function.

The reader might be alarmed that we seem to have managed to invert the
Dirac operator which is singular in the presence of a non-zero
flux. If the solution is not unique, it will differ from the one above
by a linear superposition of zero modes. This additional term does not
contribute in the calculation of the correlation function by exactly
the same arguments presented earlier for going from $g_{-+}$ to
${\tilde g}_{-+}$.

\bigskip
{\bf Appendix D}
\bigskip

The bosonic Green's function, $K_{xy}$, may be broken up into a
massive and massless part:
\begin{displaymath} K_{xy}=\langle
x|\,{1\over\Box(\Box-m^2)}\,|y\rangle=
{1\over m^2}
\langle x|\,{1\over\Box-m^2} - {1\over\Box}\,| y\rangle=\end{displaymath}
\begin{displaymath}=
{1\over m^2}\bigl(\Delta_m(x-y)+{1\over m^2 V}
-\Delta(x-y)\bigr)\end{displaymath}
as done in \cite{zahed}.

The massless part may be calculated by using the eigenmodes that span
the allowed field $\phi$ on the torus.
\begin{displaymath}\Delta(x-y)=\langle x|\,{1\over\Box}\,|y\rangle=
-\sum_{n\ne0} {\phi_n(x)
\phi^{\dagger}_n(y)\over\mu_n}\end{displaymath}
\begin{displaymath}\hbox{with}\quad \phi_n(x)={1\over{\sqrt V}}
\phi_n e^{2\pi i (n_0x^0/\beta+n_1x^1/L)}\quad\hbox{and}\quad
\phi_n\phi_m^{\dagger}=\delta_{nm}\end{displaymath}
Since $\int\phi=0$ is a necessary condition for a one to one
transformation of the fields, the value $n=(n_0,n_1)=0$ is omitted
from the summation. The eigenvalues due to the periodicity of the
torus are
\begin{displaymath}\mu_n=\Bigl({2\pi n_0\over\beta}\Bigr)^2
+ \Bigl({2\pi n_1\over
L}\Bigr)^2\end{displaymath} as found in Sachs and Wipf \cite{sw}.

The sum may be broken up into two terms: one with $n_0\ne 0$ and the
other with $n_0=0$ and $n_1\ne 0$.  Evaluating the first $n_1$ sum and
relabeling the remaining summation index as $n$,
\begin{displaymath}\Delta(x)=
-{1\over 2\pi}\sum_{n>0}\Bigl\{ {\cos(2\pi n {x^0\over\beta})\over
n}\Bigl[\coth(\pi\tau n)\cosh\bigl(2\pi n {x^1\over\beta}\bigr)-
\sinh\bigl(2\pi n {x^1\over\beta}\bigr)\Bigr]
+{\tau\over\pi}{\cos(2\pi n {x^1\over L})\over
n^2}\Bigr\}.\end{displaymath} Letting $z=(x^0+ix^1)/\beta$ and using
the identities:
\begin{displaymath}\cos(kx)\cosh(ky)=
\hbox{Re}[\cos k(x+iy)]\quad\hbox{and}\quad
\cos(kx)\sinh(ky)=\hbox{Re}[-i\sin k(x+iy)],\end{displaymath}
the first summation in the braces may be written as:
\begin{displaymath}\hbox{Re}\sum_{n>0}\Bigl[{\cos2\pi n z\over n}
(\coth(\pi\tau n)-1) + {e^{2\pi
i n z}\over n}\Bigr]=\end{displaymath}
\begin{displaymath}=\hbox{Re}\sum_{n,r>0}
\Bigl[{e^{2\pi inz}+e^{-2\pi inz}\over n}
e^{-2\pi\tau nr}\Bigr]-\ln(1-e^{2\pi iz})=\end{displaymath}
\begin{displaymath}=-\hbox{Re}\ln\Bigl\{2ie^{\pi iz} \sin(\pi
z)\prod_{r>0}(1-2e^{-2\pi\tau r}\cos(2\pi z)+e^{-4\pi\tau
r})\Bigr\}\end{displaymath} since $\coth(\pi\tau n)-1={\sum\atop{r>0}}
e^{-2\pi\tau nr}$ and $\ln(1-x)=- {\sum\atop{n>0}}{x^n\over n}$.

The second summation in the braces gives
\begin{displaymath}{\pi\tau\over6}-\pi x^1+{\pi(x^1)^2\over\tau}.
\end{displaymath}
The first two terms can be combined with the first summation to form
the product representation of $\vartheta_1$ and $\eta(i\tau)$ to
finally give
\begin{displaymath}-4\pi\Delta(x-y)=
\ln\Bigl|{\eta(i\tau)\over\vartheta_1(\pi(z_1-z_2)|i\tau)}
\Bigr|^2 + {2\pi\over V}(x^1-y^1)^2.\end{displaymath}

\end{document}